\documentclass[prb,twocolumn,superscriptaddress,longbibliography,aps,preprintnumbers]{revtex4-2}

\usepackage{graphicx}
\usepackage{bm}
\usepackage{color}
\usepackage{amsmath}
\usepackage{relsize}
\usepackage{amssymb}

\graphicspath{{Figures/}{./Figures/}{.}}

\begin{document}
\title{Higher order spin interactions mediated by the substrate}
\author{Szczepan G\l{}odzik}
\affiliation{Institute of Physics, M.\ Curie-Sk\l{}odowska University, 
20-031 Lublin, Poland}
\affiliation{Jo\v{z}ef \v{S}tefan Institute, Jamova 39, SI-1000 Ljubljana, Slovenia}
\author{Rok \v{Z}itko}
\affiliation{Jo\v{z}ef \v{S}tefan Institute, Jamova 39, SI-1000 Ljubljana, Slovenia}
\affiliation{Faculty of Mathematics and Physics, University of Ljubljana, Jadranska 19, SI-1000 Ljubljana, Slovenia}
\date{\today}

\begin{abstract}
    Since the seminal works in the fifties it has been known that a bath of non-interacting electrons mediates an interaction between local moments (such as nuclear spins or magnetic impurities) coupled to distant sites of the lattice. Recent efforts in simultaneous control of multi-qubit arrays rely on defining quantum dots in environments which likewise contain electrons capable of mediating effective interactions. It is therefore interesting to analyze systems with more than two impurities in higher orders of perturbation theory. Here we derive and discuss the significance of the effective spin interactions between four impurities in fourth order of the coupling between the impurities and the substrate, mediated by conduction electrons of a two-dimensional square lattice. This effective interaction resembles the ring exchange -- a coupling of four spins appearing in perturbative treatment of the Hubbard model. Despite it being a higher-order effect, it can dramatically influence the magnetic ordering. We show that as an effective interaction between impurity spins, the ring exchange can also compete with the more familiar two-spin interactions. 
\end{abstract}

\maketitle
\section{Introduction}
The interaction between a single magnetic moment and the Fermi sea is dominated by the rich world of Kondo physics. Many interesting phenomena arise when the number of channels or the spin of the impurity is increased~\cite{nozieres}. Furthermore, in case of multiple impurities, an exchange interaction mediated by the conduction electrons emerges~\cite{RK,K,Y}. This RKKY interaction (named after Ruderman, Kittel, Kasuya and Yosida) is long-ranged, decaying as a power-law with increasing distance between the impurities, and can be understood as Friedel oscillations of spin-polarized electrons. Because the RKKY interaction is oscillatory, the resulting magnetic ordering is heavily influenced by the concentration of the impurities. Depending on the coupling strength between the impurity spins and the itinerant electrons, either individual Kondo singlets form (for high values of the coupling), or the RKKY interaction wins and determines the magnetic order of the resulting Kondo lattice~\cite{mott,doniach}. The competition between the Kondo and RKKY interactions can become even more subtle if the spins are coupled to helical liquids~\cite{yevtushenko2018,ferrer2023}, or to nanoscopic lattices, where the spectrum of itinerant electrons becomes discrete~\cite{allerdt2015,schwabe2012}. There are systems where the lowest-order RKKY interaction can completely vanish at low temperatures~\cite{konic2023}. Continuing with the theme of increasing level of complexity in order to find interesting phenomena, considering higher order of the perturbative expansion in the coupling strength seems an exciting avenue~\cite{grytsiuk2020}, especially if lower order terms could be suppressed at the same time. A unitary transformation up to fourth order in electron hopping for the Hubbard model at half filling yields a spin Hamiltonian which includes a four-spin interaction dubbed the ring exchange~\cite{macdonald1988,dallapiazza2012}. This interaction plays an important role in magnetism of a plethora of systems~\cite{li2022,paul2020,misguich1998,lauchli2005} and has been studied in various lattices~\cite{seki2020,larsen2019}. In the context of high-$T_c$ superconductors, ring exchange (sometimes called cyclic exchange) has to be included to correctly simulate the spin wave spectra of the parent compounds~\cite{coldea2001,toader2005,katanin2002}. After this fact had been established, a large number of works turned to the study of spin ladders, where a relatively small value of the ring exchange coupling leads to spin gap closing and a quantum phase transition~\cite{gopalan1994,hikihara2003,nunner2002,brehmer1999,caponi2013}. Those efforts led to finding rich phase diagrams for both the triangular~\cite{sheng2009} and square ladders~\cite{lauchli2003}. The inclusion of the higher order spin interactions can lead to new, exotic phases~\cite{block2011,yang2010}, and recent experimental results show that the ring exchange can be the dominant interaction in an ultracold atom lattice simulating the minimal toric code~\cite{dai2017}, and that it can be engineered to entangle quartets of trapped ions~\cite{katz2023}. This prompts the question whether such a four-spin interaction between magnetic impurities can be (in analogy to the RKKY interaction) mediated by the itinerant electrons of the substrate. This is indeed the case. In this work we discuss the relative significance of the different contributions to the effective impurity-impurity interaction.  

Another class of related problems concerns the precision control of man-made systems such as arrays of singlet-triplet qubits, where accurate control of the adiabatic loading may be challenging~\cite{fedele2021}. We speculate that the four-spin interactions mediated by the environment in those types of situations might contribute to the challenges and affect the dynamics of qubit arrays, and we argue that in realistic parameter regimes it might be impossible to remove the effective higher order interactions.

The paper is organized as follows: in Sec.~\ref{sec:derivation} we derive the effective ring exchange interaction. The thermodynamic potential perturbation theory is introduced, followed by the presentation of the fourth order expansion and discussion of the parameter values. Then in Sec.~\ref{sec:results} we show the dependence of the higher order interactions on the inter-impurity distance and the chemical potential of the substrate. App.~\ref{app:sigmas} reveals how some of the terms stemming from the fourth order expansion can be reduced to two-spin interactions for spin-$1/2$ impurities. 

\section{Derivation of the effective interaction}
\label{sec:derivation}
\subsection{Thermodynamic potential perturbation theory}
We start our analysis with the usual perturbation theory setup. The full system Hamiltonian $\mathcal{H}=H_0+H_{int}$ consists of non-interacting electrons 
\begin{equation}
    H_0=-t\sum_{\langle ij\rangle\sigma}c_{i\sigma}^{\dagger}c_{j\sigma}-\mu\sum_{i\sigma} c^{\dagger}_{i\sigma}c_{i\sigma},
    \label{ham:hopping}
\end{equation}
where $\langle ij\rangle$ refers to all nearest-neighbor pairs. The interaction term consists of four magnetic impurities $\mathbf{S}_i$ Kondo coupled to the electrons of the substrate
\begin{equation}
\begin{aligned}
        H_{int}=-J(\mathbf{S}_{1}\cdot\mathbf{s}_{1}+\mathbf{S}_{2}\cdot\mathbf{s}_{2}+\mathbf{S}_{3}\cdot\mathbf{s}_{3}+\mathbf{S}_{4}\cdot\mathbf{s}_{4}),
\end{aligned}
\end{equation}
with the itinerant electrons expressed in the second quantized form $\mathbf{s}_i=\frac{1}{2}c^{\dagger}_{i\alpha}\vec{\sigma}_{\alpha\beta}c_{i\beta}$. For the purposes of deriving the effective impurity coupling mediated by itinerant electrons, we constrain the impurity spin orientations and treat the vectors $\mathbf{S}_i$ as fixed quantities (i.e., the spins are not part of the Hilbert space in the calculation that follows). The idea here is that in the limit of small $J$ the dynamics of impurity spins is slower than the dynamics of itinerant electrons, thus one can approximately decouple the two subsystems. We then employ the perturbative expansion for the thermodynamic potential of itinerant electrons $\Omega$~\cite{abrikosov}. The correction induced by the presence of impurities is expressed as 
\begin{equation}
\begin{aligned}
    \Delta\Omega&=-T \ln{\langle \mathcal{S}\rangle}\\
    \mathcal{S}&=\exp\left({-\int_0^{\beta}H_{int}(\tau)d\tau}\right),
\end{aligned}
\end{equation}
where we denote the inverse temperature $\beta=1/T$ and $H_{int}$ depends on the imaginary time $\tau=it$ through the field operators written in the interaction picture
\begin{equation}
\begin{aligned}
    c_{i\alpha}(\tau)&=e^{H_0\tau}c_{i\alpha}e^{-H_0\tau},\\
    c^{\dagger}_{i\alpha}(\tau)&=e^{H_0\tau}c^{\dagger}_{i\alpha}e^{-H_0\tau},
\end{aligned}
\end{equation}
where we keep the dagger symbol even though $c^{\dagger}_{i\alpha}(\tau)$ is not the hermitian conjugate of $c_{i\alpha}(\tau)$.
This approach is particularly convenient because one can show that $\Delta\Omega$ can be expressed as a series 
\begin{equation}
    \Delta\Omega=-T \left(\langle \mathcal{S}\rangle_{con}-1\right)=-T[\Xi_1+\Xi_2+\ldots],
\end{equation}
where the correction at order $n$ 
\begin{equation}
    \Xi_n=\dfrac{(-1)^n}{n!}\int d\tau_1\ldots d\tau_n\langle T_{\tau}(H_{int}(\tau_1)\ldots H_{int}(\tau_n)) \rangle_{con}
    \label{correction_n}
\end{equation}
is evaluated by only considering the connected diagrams when applying Wick's theorem to the expectation values $\langle T_{\tau}(H_{int}(\tau_1)\ldots H_{int}(\tau_n)) \rangle$. $T_{\tau}$ is the time ordering operator. Such reasoning was previously applied to the detailed study of the RKKY interaction between impurity moments in graphene~\cite{saremi2007,kogan2011}. Another advantage of this approach is that the calculation results in the effective interaction expressed in terms of products of Matsubara Green's functions, which by analytic continuation can be related to the retarded Green's functions of electrons in the underlying lattice. In other words, the goal of this approach is to integrate out the lattice electrons in order to obtain an effective spin-only Hamiltonian for the impurities. Since we assume that there are no external fields all $\Xi_n$ with $n$ odd vanish due to time reversal symmetry. In second order, we recover the RKKY interaction between all six pairs of impurities and a constant energy shift.

\subsection{Fourth order expansion}
The fourth order $\Xi_4$ is computed using a computer algebra system to manage the complex expressions~\cite{sneg}. The general form of those is 
\newcommand{\bsigma}{\boldsymbol{\sigma}}
\begin{widetext}
    \begin{equation}
        \begin{aligned}
            \tfrac{1}{16}J^4\sum_{\substack{\alpha\beta\gamma\delta\\ \varepsilon\zeta\eta\theta}}\mathbf{S}_1\cdot\bsigma_{\alpha\beta} \, \mathbf{S}_2\cdot\bsigma_{\gamma\delta} \, \mathbf{S}_3\cdot\bsigma_{\varepsilon\zeta} \, \mathbf{S}_4\cdot\bsigma_{\eta\theta}%
\int d\tau_1 d\tau_2 d\tau_3 d\tau_4  \langle c_{1\alpha}^{\dagger}(\tau_1)c_{1\beta}(\tau_1) c_{2\gamma}^{\dagger}(\tau_2)c_{2\delta}(\tau_2) c_{3\varepsilon}^{\dagger}(\tau_3)c_{3\zeta}(\tau_3) c_{4\eta}^{\dagger}(\tau_4)c_{4\theta}(\tau_4)\rangle
        \end{aligned}
        \label{typical_term}
    \end{equation}
\end{widetext}

There are six topologically equivalent connected diagrams \footnote{Unlike in the usual perturbation theory, when considering the $\mathcal{S}$ matrix for the thermodynamic potential the number of topologically equivalent diagrams is $(n-1)!$ and it does not fully cancel the $1/n!$ prefactor in the expression for $\Xi_n$ Eq.~\ref{correction_n}.}. In the absence of spin-orbit coupling only terms which conserve spin need to be retained. We define the Matsubara Green's functions in imaginary time
\begin{equation}
    -\mathcal{G}_{ij}(\tau_2-\tau_1)=\langle T_{\tau} c_{i}(\tau_1)c_j^{\dagger}(\tau_2)\rangle,
\end{equation}
omitting the spin labels in the electron creation and annihilation operators because the ground state in unpolarized, leading to $\langle c_{i\uparrow}(\tau_1)c^{\dagger}_{j\uparrow}(\tau_2) \rangle =\langle c_{i\downarrow}(\tau_1)c^{\dagger}_{j\downarrow}(\tau_2) \rangle $. To complete the analysis, we expand the Matsubara Green's functions in a Fourier series
\begin{equation}
    \mathcal{G}_{ij}(\tau)=\frac{1}{\beta}\sum_ne^{-i\omega_n\tau}G_{ij}(i\omega_n),
\end{equation}
with discrete Matsubara frequencies $\omega_n=(2n+1)\pi/\beta$, to transform the convolution of four $\mathcal{G}_{ij}(\tau)$ functions into a product. Recognizing the Fourier representation of the $\delta$ functions $\frac{1}{\beta}\int d\tau \exp\left(-i\tau(\omega_n-\omega_m)\right)=\delta_{n,m}$ and taking into account the real-space symmetry of the Green's functions $G_{ij}(i\omega_n)=G_{ji}(i\omega_n)$ we arrive at the final expression, which consists of six general types of contributions to the effective interaction between the impurity spins. 

After the electrons are ``integrated out'' to find the expression for $\Delta\Omega$ for fixed directions of vectors $\mathbf{S}_i$, these are promoted to quantum mechanical operators and the expression is reinterpreted as an effective spin Hamiltonian for the impurities. The approximation is controlled by the timescales of the resulting effective spin-spin couplings (both RKKY and higher-order). 

Before stating the final result, we discuss some of those contributions, as further simplifications arise by assuming that the impurity spins are spin-$1/2$ objects with no internal orbital structure. The first rather obvious simplification is the constant energy shift coming from expansion terms of the type $(S_i^{x,y,z})^2(S^{x,y,z}_j)^2=1/16$ and $(S^{x,y,z}_i)^4=1/16$. Another unsurprising result is the emergence of RKKY-like two-impurity interaction in the fourth order in $J$. The third simplification are the effective biquadratic interactions $(\mathbf{S}_i\cdot\mathbf{S}_j)^2$, which reduce to two-spin interactions and further contributions to the constant shift via $(\mathbf{S}_i\cdot\mathbf{S}_j)^2=\frac{3}{16}-\frac{1}{2}\mathbf{S}_i\cdot\mathbf{S}_j$ (see Appendix~\ref{app:sigmas}). The last type of terms to be discussed at the present stage are the three-spin interactions. There appear twelve terms with the structure $(\mathbf{S}_i\cdot\mathbf{S}_j)(\mathbf{S}_i\cdot\mathbf{S}_k)$ that can again be reduced to two-spin interactions $\mathbf{S}_j\cdot\mathbf{S}_k$ (see Appendix~\ref{app:sigmas}). With those remarks in place, we can now state the full expression for $\Xi_4$:
\begin{widetext}
    \begin{equation}
        \begin{aligned}           \Xi_4=E^{(4)}+\sum_{i\neq j}\chi_{ij}\mathbf{S}_i\cdot \mathbf{S}_j+\chi_{ring}
        \end{aligned}
        \label{eq:fourthorder}
    \end{equation}
    with the constant shift 
    \begin{equation}
        E^{(4)}=\tfrac{9}{32}J^4\mathlarger{\sum_{n}}\Bigl(G^4_{00}(i\omega_n)+\sum_{i\neq j}\bigl[G^2_{00}(i\omega_n)G^2_{ij}(i\omega_n)-\tfrac{2}{3}G^4_{ij}(i\omega_n) \bigr]\Bigr),
    \end{equation}
    the fourth order correction to the two-spin susceptibility
    \begin{equation}
        \chi_{ij}=\tfrac{1}{16}J^4\mathlarger{\sum_n}\Bigl(3G^2_{00}(i\omega_n)G^2_{ij}(i\omega_n)-G^4_{ij}(i\omega_n)+\sum_{k\neq\lbrace i,j\rbrace}\bigl[-\tfrac{1}{2}G^2_{ik}(i\omega_n)G^2_{jk}(i\omega_n)+3G_{00}(i\omega_n)G_{ij}(i\omega_n)G_{ik}(i\omega_n)G_{jk}(i\omega_n)\bigr]\Bigr),
        \label{eq:twospin4}
    \end{equation}
and the ring exchange term which consists of three contributions
\begin{equation}
    \chi_{ring} = \chi_1 + \chi_2 + \chi_3,
\end{equation}
where
\begin{equation}
\begin{aligned}
    \chi_1=\tfrac{1}{4}J^4\mathlarger{\sum}\limits_{n}&G_{12}(i\omega_n)G_{23}(i\omega_n)G_{34}(i\omega_n)G_{41}(i\omega_n)\bigl[\phantom{-}(\mathbf{S}_{1}\cdot\mathbf{S}_{2})(\mathbf{S}_{3}\cdot\mathbf{S}_{4})+(\mathbf{S}_{1}\cdot\mathbf{S}_{4})(\mathbf{S}_{2}\cdot\mathbf{S}_{3})-(\mathbf{S}_{1}\cdot\mathbf{S}_{3})(\mathbf{S}_{2}\cdot\mathbf{S}_{4}) \bigr],\\
    \chi_2=\tfrac{1}{4}J^4\mathlarger{\sum}\limits_{n}&G_{12}(i\omega_n)G_{24}(i\omega_n)G_{43}(i\omega_n)G_{31}(i\omega_n)\bigl[\phantom{-} (\mathbf{S}_{1}\cdot\mathbf{S}_{2})(\mathbf{S}_{3}\cdot\mathbf{S}_{4})-(\mathbf{S}_{1}\cdot\mathbf{S}_{4})(\mathbf{S}_{2}\cdot\mathbf{S}_{3})+(\mathbf{S}_{1}\cdot\mathbf{S}_{3})(\mathbf{S}_{2}\cdot\mathbf{S}_{4}) \bigr],\\
    \chi_3=\tfrac{1}{4}J^4\mathlarger{\sum}\limits_{n}&G_{14}(i\omega_n)G_{42}(i\omega_n)G_{23}(i\omega_n)G_{31}(i\omega_n)\bigl[- (\mathbf{S}_{1}\cdot\mathbf{S}_{2})(\mathbf{S}_{3}\cdot\mathbf{S}_{4})+(\mathbf{S}_{1}\cdot\mathbf{S}_{4})(\mathbf{S}_{2}\cdot\mathbf{S}_{3})+(\mathbf{S}_{1}\cdot\mathbf{S}_{3})(\mathbf{S}_{2}\cdot\mathbf{S}_{4}) \bigr].
\end{aligned}
\label{eq:ring}
\end{equation}
\end{widetext}
The ring exchange displays a richer structure compared to the form familiar from the Hubbard model expansion, because there are no restrictions as to the placement of the impurities and because the interaction between them is mediated by the itinerant electrons which do not participate in the formation of local moments (in contrast to the direct nearest-neighbor hopping in the Hubbard model between lattice sites which themselves host the local moments). The three terms differ in the structure of the products of Matsubara Green's functions reflecting the corresponding connected Feynman diagrams. 

Having obtained the form of the effective ring exchange of four spins, we can start exploring its significance and evolution with respect to the distance between the impurities, or the chemical potential of the underlying lattice. We can employ the known expressions for the lattice Green's function for different Bravais lattices~\cite{katsura1971,horiguchi1971} by performing the analytic continuation $    G_{ij}^R(\omega)=\lim_{i\omega_n\rightarrow \omega+i0^+}G_{ij}(i\omega_n)$. We consider the case of square lattice. 
The definition based on a recurrence relation introduced by Morita was found to be particularly useful in this case~\cite{morita1971}.
\subsection{Discussion of parameter values}
Clearly, if $J$ is tiny, the fourth order corrections will be negligible. 
We are mostly interested in the case where $J$ is small to moderate, so that the perturbation theory makes sense if truncated to fourth order. We will be particularly interested in the possible occurrences of the situation where the details of the system (spatial arrangement of the impurities or the band filling) lead to the fourth order interactions becoming dominant, while sixth order corrections are small enough to be of no concern.

To pick a sensible value for numerical evaluation, we consider impurities that emerge from the Anderson impurity model \cite{anderson} that are well in the magnetic part of the phase diagram. The Schrieffer-Wolff transformation shows that Anderson and Kondo models are related at half filling by $J_{SW}=-8|V|^2/U$, where $V$ is the matrix element of the overlap between the localized states of the impurity and itinerant electrons, and $U$ is the Coulomb repulsion of the impurity electrons~\cite{schriefferwolff}. The minimum value of $U$ for which the impurity behaves like a localized magnetic moment is $U=\pi^2\rho|V|^2$. Assuming a structureless density of states, choosing hopping in Eq.~\ref{ham:hopping} to be $t=0.5$ gives the bandwidth of the square lattice $2D=4$. Inserting the Anderson limit into the effective coupling formula gives the highest value of the physically acceptable coupling: $J=32/\pi^2 \approx 3.25$. 
Based on these considerations, in the following  we use $J=D/4=0.5$ which is physically realistic and represents the generic intermediate-coupling situation.

If the underlying lattice would realise the so called Kondo box, i.e., a situation where the itinerant electrons' spectrum is discrete, large values of $J$ would put the system in a regime where it is actually the Kondo interaction that wins, and the impurities are uncorrelated~\cite{schwabe2012,allerdt2015}. Even more dramatically, an inverse effect, where the Kondo singlets mediate the interaction between local moments formed by conduction electrons may arise~\cite{schwabe2013}. 

Since we do not aim to study the temperature dependence of the problem, we focus on low temperatures and set $\beta=10^4$. 

\section{Ring exchange for different impurity configurations}
\label{sec:results}

\begin{figure}
    \centering
    \includegraphics[width=\linewidth]{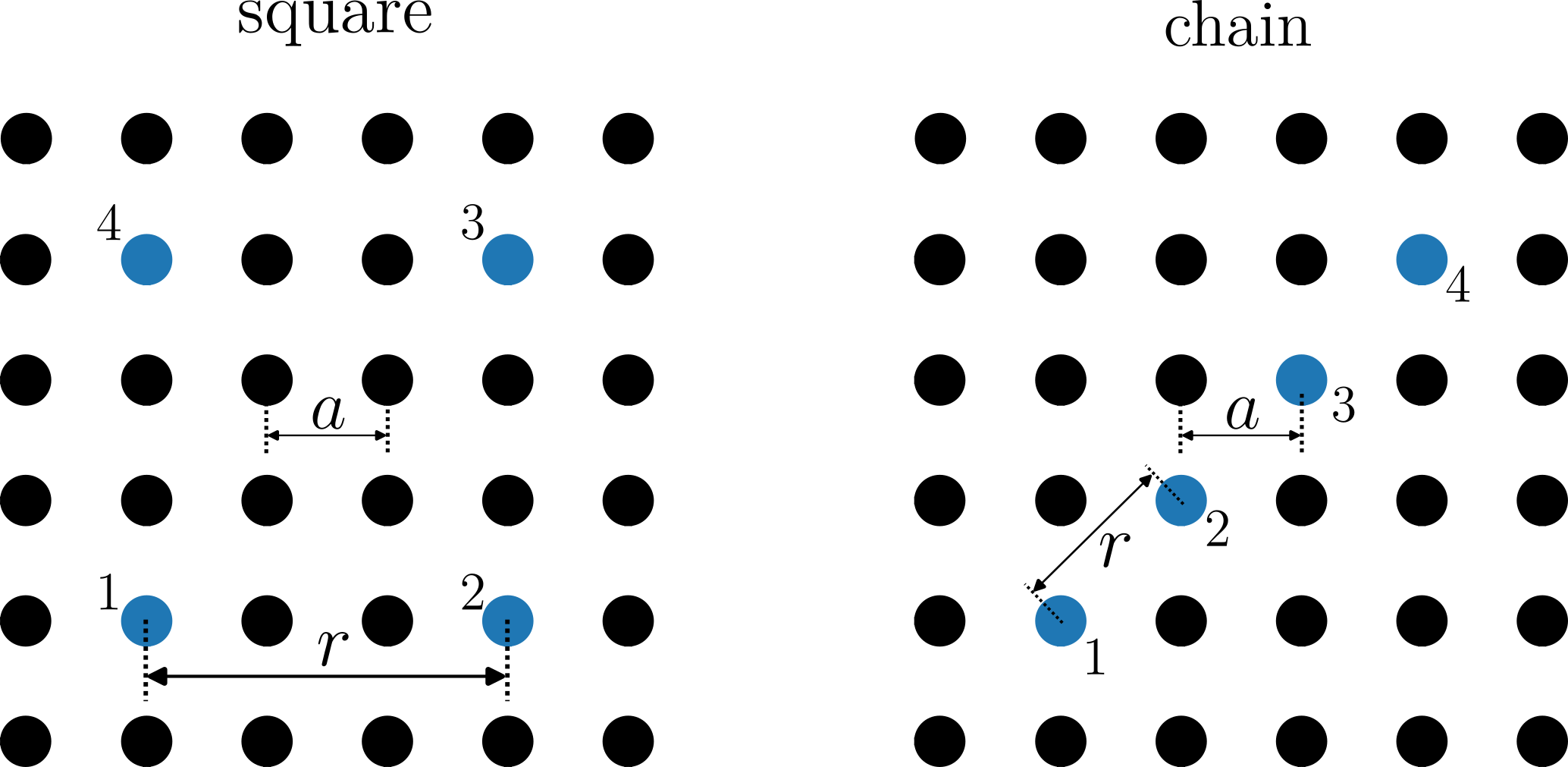}
    \caption{Lattice and geometries of the impurity placement. Black dots represent the underlying square lattice, while the blue dots are the sites with the attached impurities. The lattice constant $a$ and the shortest distance between impurities $r$ are shown. We refer to these two arrangements as 'square' and 'chain', respectively. }
    \label{fig:lattice}
\end{figure}
We now discuss the variation of the effective ring exchange with the distance between the impurities. 
We first consider the case of half filling with one conduction band electron per site.
Owing to the particle-hole symmetry that holds at half filling on the bipartite square lattice, the second-order RKKY interaction is (anti)ferromagnetic if the impurities are coupled to sites belonging to the (opposite) same sublattice~\cite{kogan2011,saremi2007}. 
All three contributions to the ring exchange are also oscillatory as a function of increasing distance between the impurities in the square arrangement (as in the left panel in Fig.~\ref{fig:lattice}).

\begin{figure}
    \centering
    \includegraphics[width=\linewidth]{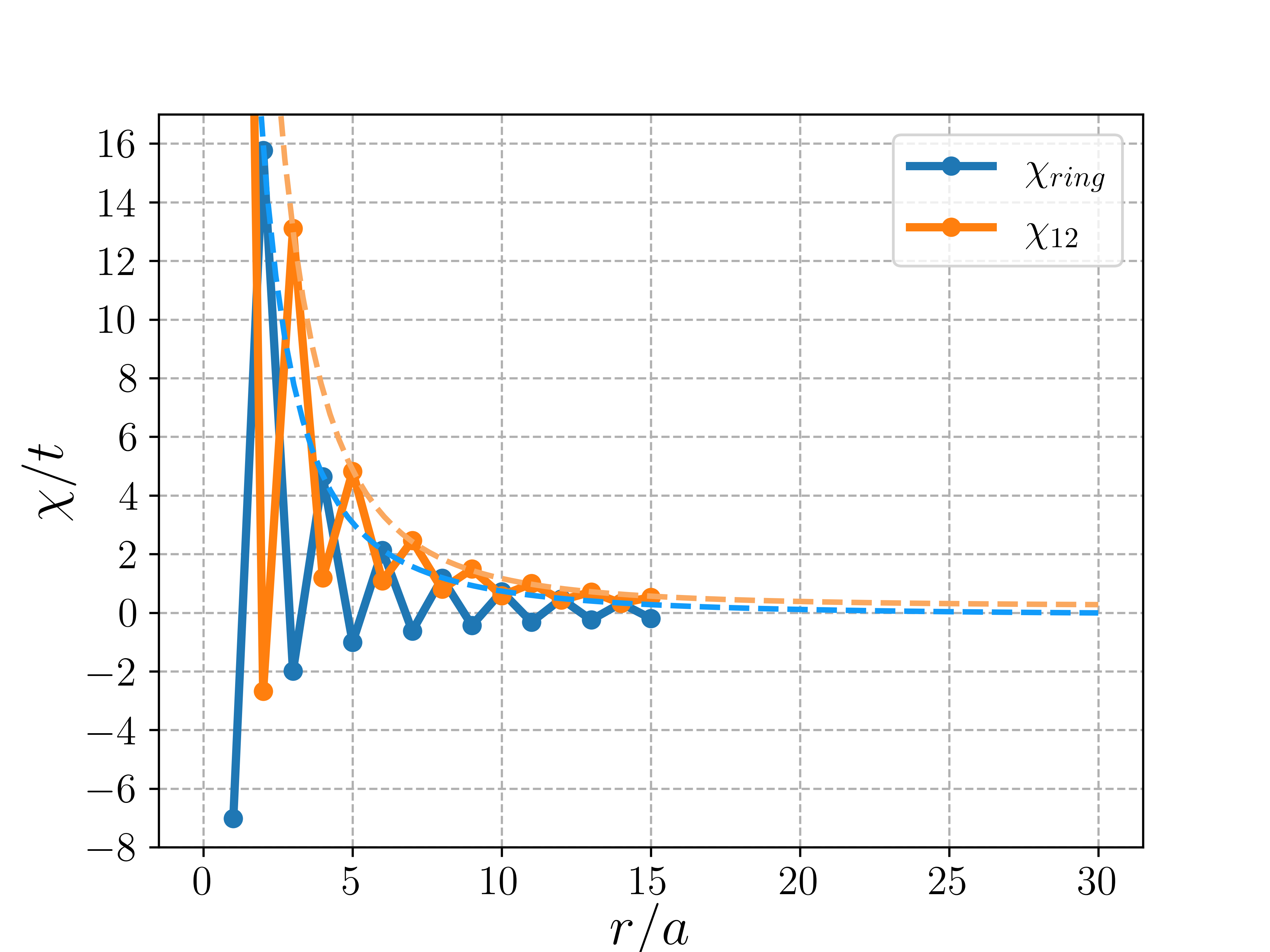}
    \caption{Sum of the effective ring exchange contributions $\chi_{ring}$ and sum of second and fourth order contributions to the two-spin interactions $\chi_{12}$ with increasing $r$ in the square arrangement for $\mu=0$. The dashed lines represent the power law fit passing through the (even) odd-$r/a$ data points for ($\chi_{ring}$) $\chi_{12}$.}
    \label{fig:square_dist}
\end{figure}

To compare the strength of the ring exchange to the two-spin RKKY, we plot in Fig.~\ref{fig:square_dist} the sum of all three contributions to ring exchange alongside the sum of both second-order $\chi^{RKKY}_{12}=\frac{1}{2}J^2\sum_{n}G^2_{12}(i\omega_n)$ and fourth-order (Eq.~\ref{eq:twospin4}) parts of the two-spin interaction between the impurities $1$ and $2$. 
On square lattice $G_{12}=G_{23}=G_{34}=G_{14}$ and $G_{13}=G_{24}$. For the square arrangement this leads to two of the three ring exchange contributions be equal, $\chi_2=\chi_3$.  

Both curves in Fig.~\ref{fig:square_dist} are oscillatory, but the two-spin interaction is strictly positive (antiferromagnetic) for $r>2$. This tendency for antiferromagnetism can be explained by the fact that the fourth order contribution to the two-spin interactions contains terms such as $G_{13}(i\omega_n)$ in the expression for $\chi_{12}$ and impurities $1$ and $3$ reside on the same sublattice, regardless of $r$ being even or odd. As we remarked earlier, particle-hole symmetry leads to antiferromagnetic ordering when the impurities couple to the same sublattice. The ring exchange is particularly large for even distances when all four impurities are coupled to the same sublattice. We fit the envelopes of the ring exchange and two-spin interactions and find that they decay as a power law, with effective exponents $-2.06$ and $-2.83$ respectively. We mark that for this choice of $J$, the effective four-spin interactions are comparable to two-spin coupling and 
therefore relevant, in particular for the case of two-spin interactions between impurities which reside on the side of the square. 

As mentioned before, regardless of the value of $r$, the pairs of impurities $1$-$3$ and $2$-$4$ couple to the same sublattice, thus their RKKY interaction is ferromagnetic. The fourth-order contribution to the interaction in those pairs contains Green's functions for other pairs, but in this case it is always antiferromagnetic. Because the fourth-order correction is obviously smaller than the second-order RKKY term, the net effect is a ferromagnetic interaction, slightly reduced in strength. 

\begin{figure}
    \centering
    \includegraphics[width=\linewidth]{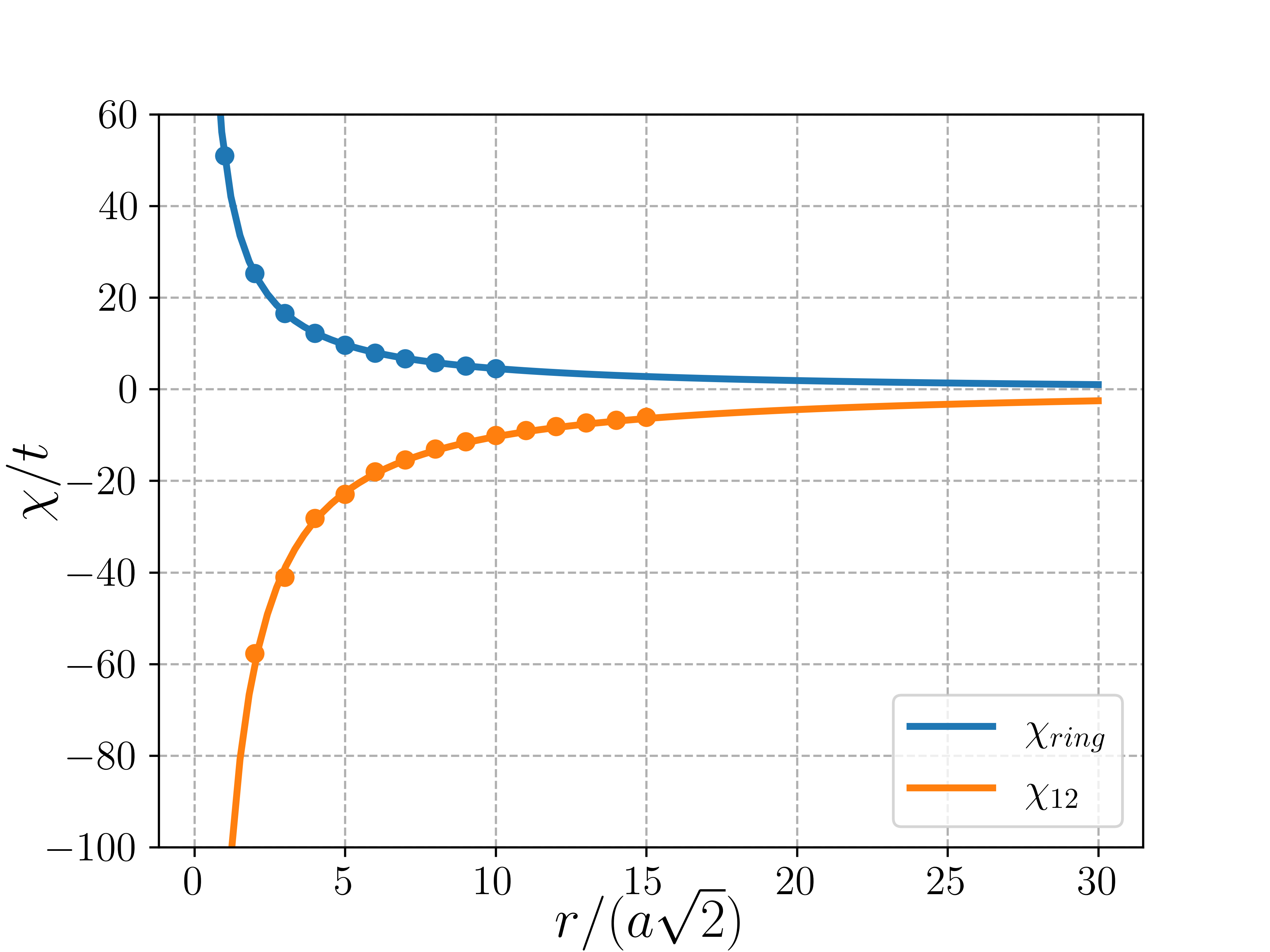}
    \caption{Sum of the effective ring exchange contributions $\chi_{ring}$ and sum of second and fourth order contributions to the two-spin interactions $\chi_{12}$ with increasing $r$ in the chain arrangement for $\mu=0$. The points are numerical results and the lines are a power-law fit. }
    \label{fig:chain_dist}
\end{figure}

Having observed that when all four impurities are placed on the same sublattice, the structure of Green's functions leads to particularly large values of the ring exchange, we now focus on such configurations. Impurities forming a chain (Fig.~\ref{fig:lattice}) is an example of an arrangement which maximizes the relative strength of the ring exchange. We again compare the total ring exchange with the total two-spin interaction as a function of $r$ in Fig.~\ref{fig:chain_dist}. For this configuration, the ring exchange does not oscillate and it decays in a similar way as $\chi_{12}$. In fact, we find that both interactions decay as $r^{-1}$. 

\subsection{Order and frustration}

\begin{figure}
    \centering
    \includegraphics[width=\linewidth]{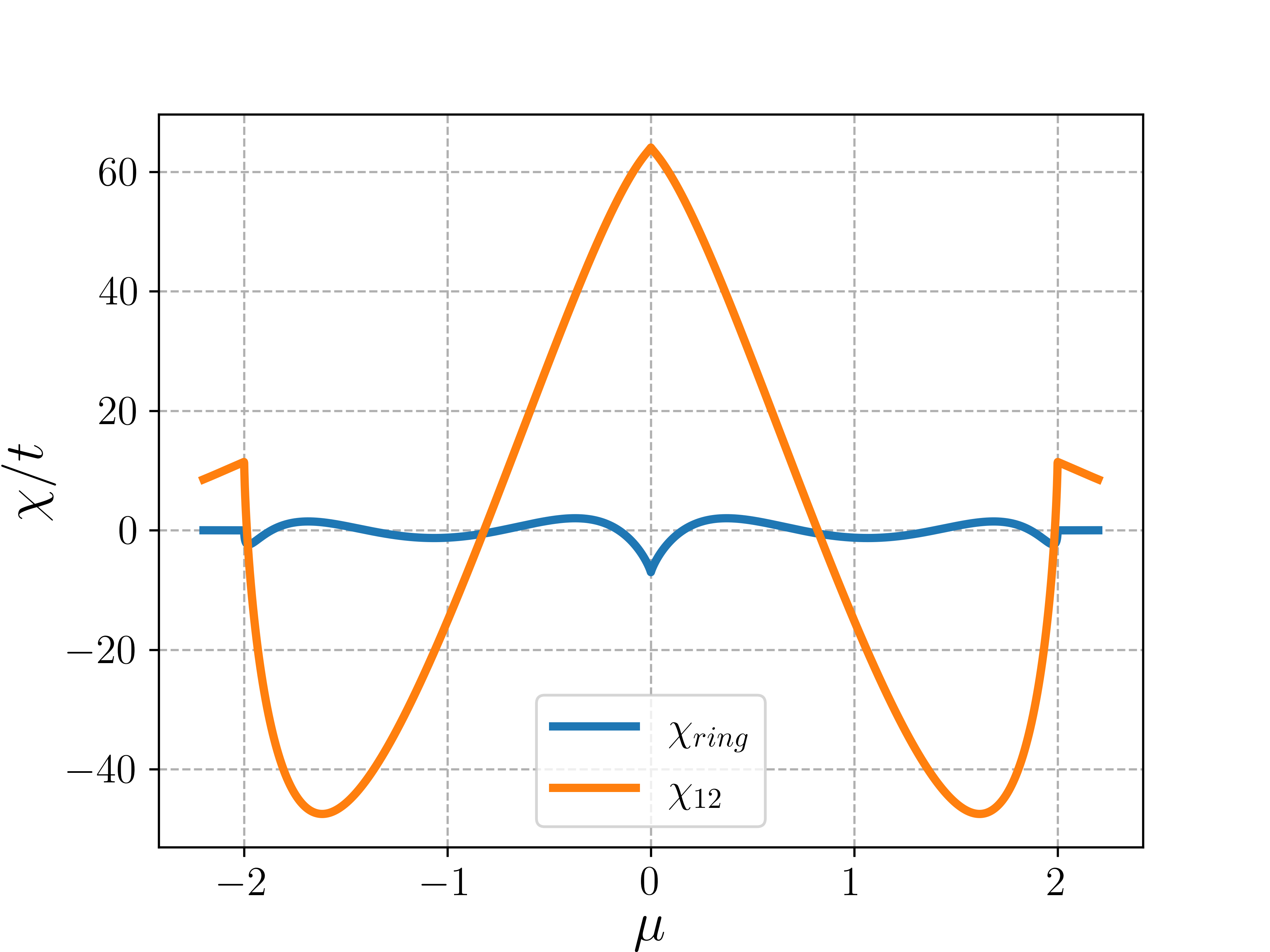}\\ \vspace{-20pt}
    \includegraphics[width=\linewidth]{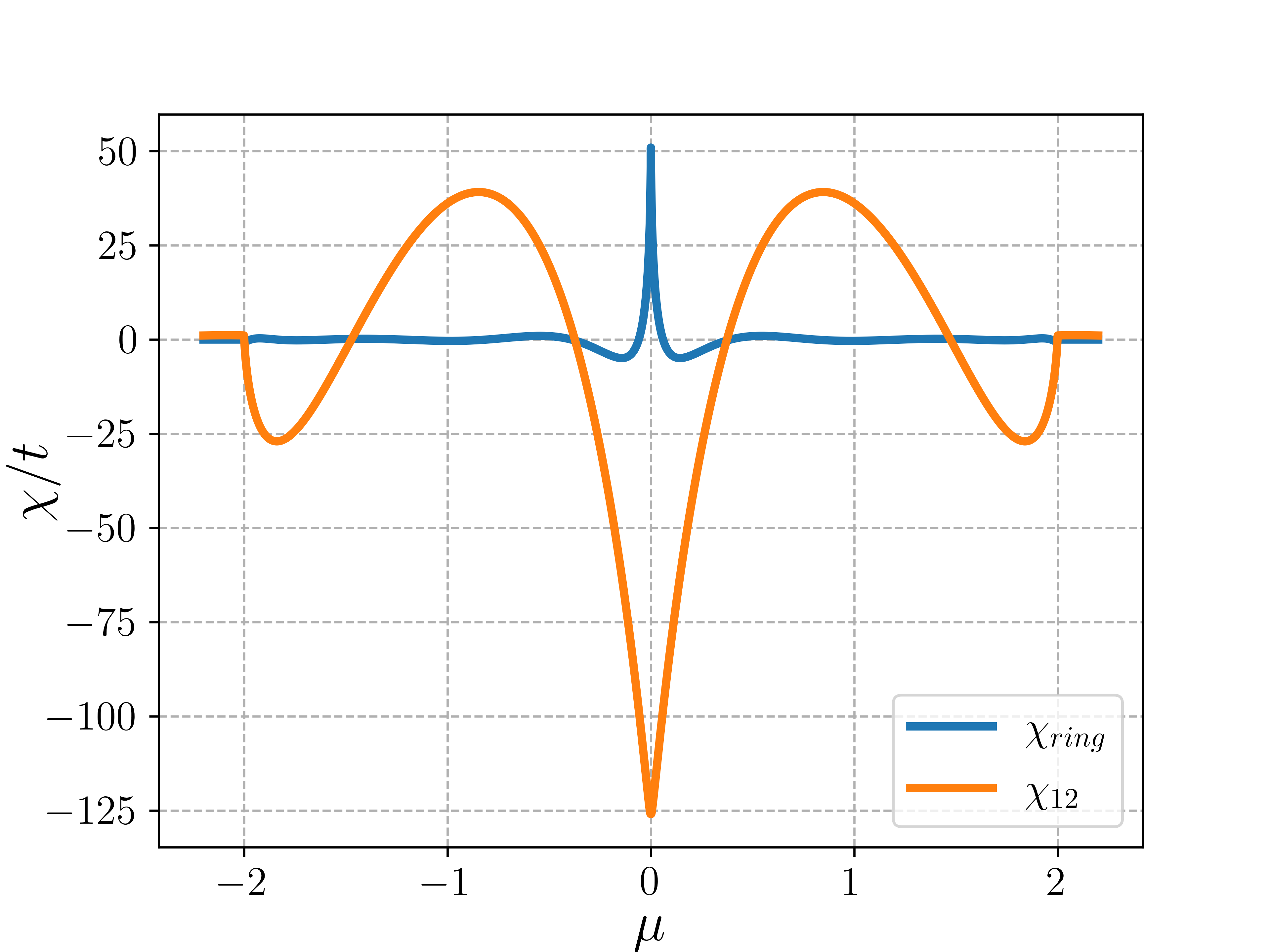}
    
    \caption{Sum of the effective ring exchange contributions $\chi_{ring}$ and sum of second and fourth order contributions to the two-spin interactions $\chi_{12}$  with respect to the chemical potential $\mu$ for square arrangement with $r=a$ (top) and chain with $r=a\sqrt{2}$ (bottom).}
    \label{fig:vs_mu}
\end{figure}

Now we turn to the dependence on the band filling. The variation of the ring exchange and two-spin interactions with the chemical potential is presented in Fig.~\ref{fig:vs_mu}. Both are maximal in amplitude at half filling, $\mu=0$. Furthermore, the strength of the ring exchange at that point is relatively the highest compared to the two-spin interactions. The presence of four impurities and the changing sign of the interactions with doping mean that one can look for situations where the alignment of impurity spins due to the effective interaction can be frustrated or stabilized.

Let us consider the square arrangement as an example. If the two-point interaction between the impurities $1$ and $2$ (and other pairs which reside on the sides of the square) is ferromagnetic ($\chi_{12}=\chi_{14}<0$), and simultaneously, the interaction between impurities $1$ and $3$ (on the diagonal) is antiferromagnetic, there is a degree of frustration, as spins $1$ and $3$ favor anti-alignment, while both favor alignment with spin $2$. This situation would correspond to $\chi_{12}+\chi_{14}+\chi_{13}=0$ being satisfied. If however $\chi_{12}=\chi_{14}>0$, meaning that the impurities tend to anti-align with their neighbors on the side of the square, while $\chi_{13} < 0$ leading to ferromagnetic interaction between impurities $1$ and $3$, the square would tend to order in a Néel state. 

Two further possibilities arise when $\chi_{12}+\chi_{14}-\chi_{13}=0$, that is, when the diagonal contribution in the square has the same sign as the contributions on the sides. If now $\chi_{12}$ is (anti)ferromagnetic, the system is (frustrated) ferromagnetically ordered. 

For the intermediate value of $J$ which we consider, we did not find any value of chemical potential which would lead to the stable (Néel of ferromagnetic order) situation described above. We did however manage to find the two types of frustration in the square arrangement, i.e. $\mu=\pm1.03$ when $\chi_{12}<0$ and $\chi_{13}>0$, as well as $\mu=\pm0.635$ when $\chi_{12}>0$ and $\chi_{13}>0$. 

\begin{figure}
    \centering
    \includegraphics[width=\linewidth]{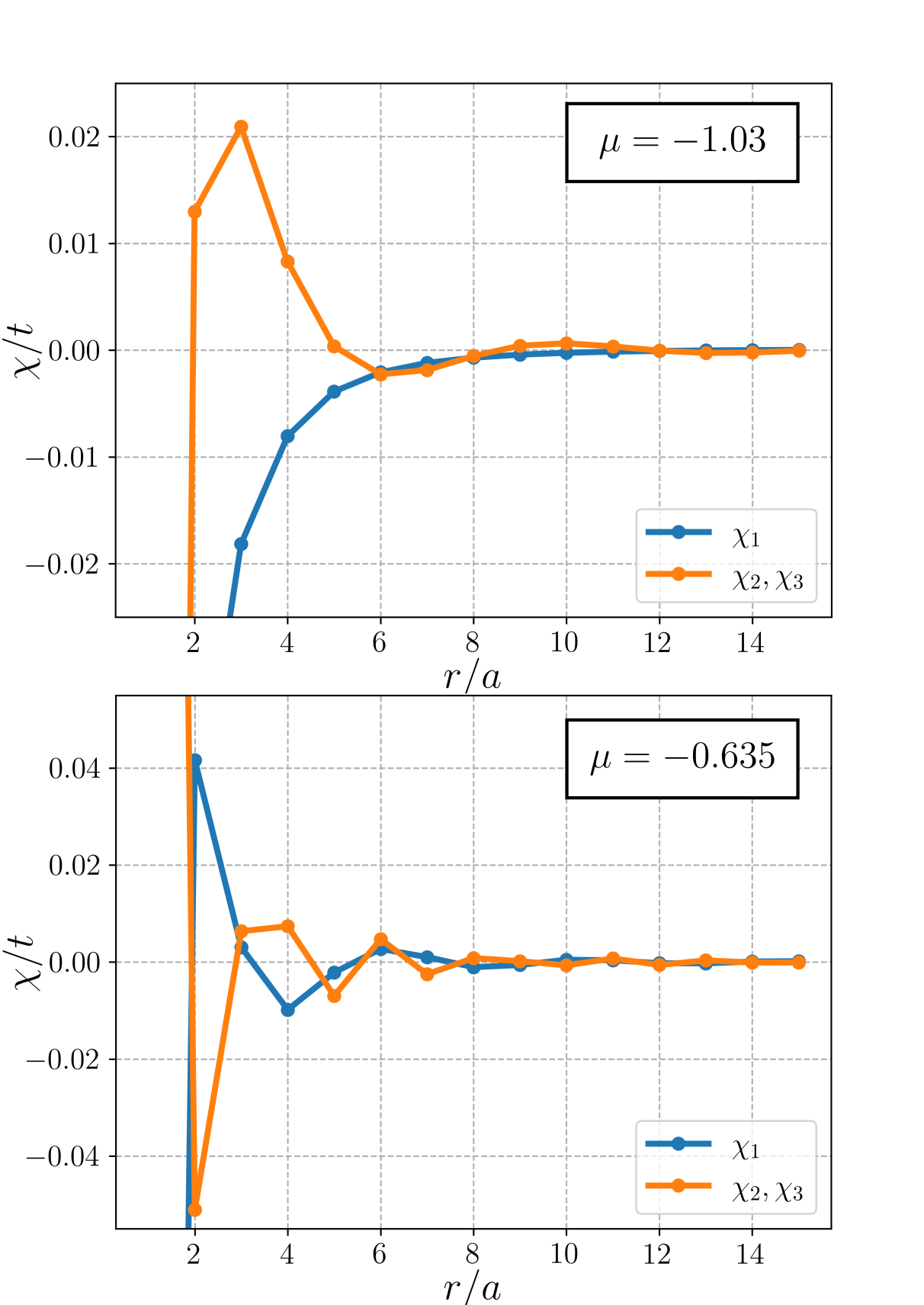}
    \caption{Ring exchange with respect to the size of the square arrangement $r$ for two different chemical potential values, corresponding to the frustration points described in the main text.}
    \label{fig:square_frust}
\end{figure}

The dependence of the ring-exchange terms $\chi_1$ and $\chi_2=\chi_3$ on $r$ is presented in Fig.~\ref{fig:square_frust} for the two values of $\mu$ just identified. For $\mu=-1.03$ different ring exchange contributions behave in a different way. While $\chi_1$ is always negative, $\chi_2$ and $\chi_3$ oscillate, with a much larger period than in the particle-hole symmetric point. The inspection of the $y$-axis reveals that the contribution of the four-spin interactions is much smaller with finite doping. The same holds for $\mu=-0.635$; in this case all ring exchange terms oscillate with some short-distance period. In both cases the ring exchange quickly decays to very small values. 

Another important point is that while we present here the $r$ dependence of the four-spin interactions, it is only at $r=1$ where the conditions $\chi_{12}+\chi_{14}\pm\chi_{13}=0$ are met for the specific chemical potentials used. There is no realistic way of modulating the chemical potential in real space so as to fulfill those conditions regardless of the distance between the impurities. In general, both stability and frustration require fine tuning in such a complicated system with many types of competing interactions, which strongly depend on the value of the chemical potential. We have examined different arrangements, for example a square arrangement rotated by $45^{\circ}$ with respect to the lattice vectors, or chains aligned with the lattice vectors. All those other types of arrangements did not produce dramatically different results, apart from the ring exchange being relatively weaker when compared to the two-spin interactions.

\section{Conclusions}
We have shown that four Kondo impurities in a host material are indirectly coupled through an effective four-spin interaction mediated by the itinerant electrons of the underlying lattice. By deriving the full fourth-order perturbative correction to the thermodynamic potential, we have determined that this substrate-mediated ring exchange can be comparable in magnitude to the two-spin interactions, and that for realistic intermediate coupling strength parameters it is difficult to tune it away. Different arrangements of impurities attached to the sites of a square lattice lead to different real-space characteristics of the effective interaction, strongly depending on the chemical potential and the impurity locations (in particular with respect to sublattices). Despite the relative strength of the ring exchange significantly decreasing with doping away from half-filling, we have found the possibility to tune the system to frustration or stability points, which could perhaps be implemented in artificial lattices.

A possible extension of the present study is considering other lattices, as for short inter-impurity distances it is possible to calculate the appropriate Green's functions numerically. Triangular lattice presents itself as an immediate candidate for an extension. Additionally, one can expect that when the time reversal symmetry is broken, effective chiral spin interactions of the type $\mathbf{S}_i\cdot(\mathbf{S}_j\times\mathbf{S}_k)$ can be mediated by the lattice. 

\acknowledgments
This work was supported by the Slovenian Research and Innovation Agency (ARIS) under grants J1-3008 and P1-0416. S. G. thanks Luka Pave\v{s}i\v{c} for illuminating discussions. Raw data for the plots and the full expression of the fourth order perturbative expansion can be found on Zenodo~\cite{zenodo}.

\appendix

\section{Impurity spin expressions}\label{app:sigmas}
Here we present the logic behind the reduction of some terms appearing in the fourth order of perturbation theory, which is applicable when the impurities are spin-$1/2$ particles. In that case one can express the spin operators via the Pauli matrices $2S^{x,y,z}=\sigma^{x,y,z}$, and use the familiar identity $\sigma^a\sigma^b=\delta_{ab}+i\varepsilon_{abc}$ to write:
\begin{equation}
\begin{aligned}
    (\mathbf{S}_i\mathbf{S}_j)^2&=S^a_iS^a_jS^b_iS^b_j=(\tfrac{1}{4}\delta_{ab}+\tfrac{i}{2}\varepsilon_{abc}S^c_i)(\tfrac{1}{4}\delta_{ab}+\tfrac{i}{2}\varepsilon_{abc}S^c_j)\\
    &=\tfrac{3}{16}-\tfrac{1}{2}\mathbf{S}_i\mathbf{S}_j.
\end{aligned}
\end{equation}
The product comprising three spins can be treated in a similar way:
\begin{equation}
    \begin{aligned}
        (\mathbf{S}_i\mathbf{S}_j)(\mathbf{S}_i\mathbf{S}_k)&=S^a_iS^a_jS^b_iS^b_k=S^a_jS^b_k(\tfrac{1}{4}\delta_{ab}+\tfrac{i}{2}\varepsilon_{abc}S^c_i)\\
&=\tfrac{1}{4}\mathbf{S}_j\mathbf{S}_k+\tfrac{i}{2}\varepsilon_{abc}S_i^aS_k^bS_i^c.
    \end{aligned}
\end{equation}
Because the same number of terms in opposite order appear, i.e. $(\mathbf{S}_i\mathbf{S}_k)(\mathbf{S}_i\mathbf{S}_j)$, the imaginary terms cancel. 

\clearpage

\bibliography{bibliography}

\end{document}